\documentclass[fleqn,usenatbib]{mnras}

\usepackage{newtxtext,newtxmath}

\usepackage[T1]{fontenc}

\DeclareRobustCommand{\VAN}[3]{#2}
\let\VANthebibliography\thebibliography
\def\thebibliography{\DeclareRobustCommand{\VAN}[3]{##3}\VANthebibliography}

\usepackage{graphicx}	
\usepackage{amsmath}	
\usepackage{tabularray}
\usepackage{lscape}
\usepackage[flushleft]{threeparttable}
\usepackage{dirtytalk}

\title[Long-term ETVs in white dwarf binaries]{Long-term eclipse time variations in white dwarf binaries}

\author[A. Yates et al.]{
Amalie Yates,$^{1}$\thanks{E-mail: alyates2@sheffield.ac.uk}
S. G. Parsons,$^{1}$
A. J. Brown,$^{2}$
N. Castro Segura,$^{3}$
V. S. Dhillon,$^{1,4}$
M. J. Dyer,$^{1,5}$
\newauthor
J. A. Garbutt,$^{1}$
M. J. Green,$^{6,7}$
D. Jarvis,$^{1}$
M. R. Kennedy,$^{8}$
P. Kerry,$^{1}$
D. Kilkenny,$^{9}$
S. P. Littlefair,$^{1}$
\newauthor
J. McCormac,$^{3}$
J. Munday,$^{3}$
I. Pelisoli,$^{3}$
E. Pike,$^{1}$
D. I. Sahman$^{1}$
\\
$^{1}$Astrophysics Research Cluster, School of Mathematical and Physical Sciences, University of Sheffield, Sheffield, S3 7RH, UK\\
$^{2}$Hamburger Sternwarte, University of Hamburg, Gojenbergsweg 112, 21029 Hamburg, Germany\\
$^{3}$Department of Physics, University of Warwick, Gibbet Hill Road, Coventry, CV4 7AL, UK\\
$^{4}$Instituto de Astrofísica de Canarias, E-38205 La Laguna, Tenerife, Spain\\
$^{5}$Research Software Engineering, University of Sheffield, Sheffield, S1 4DP, United Kingdom\\
$^{6}$Homer L. Dodge Department of Physics and Astronomy, University of Oklahoma, 440 W. Brooks Street, Norman, OK 73019, USA\\
$^{7}$JILA, University of Colorado and National Institute of Standards and Technology, 440 UCB, Boulder, CO 80309-0440, USA\\
$^{8}$School of Physics, University College Cork, Cork, T12 K8AF, Ireland\\
$^{9}$Department of Physics \& Astronomy, University of the Western Cape, Bellville 7535, South Africa\\
}

\date{Accepted XXX. Received YYY; in original form ZZZ}

\pubyear{\the\year{}}

\begin{document}
\label{firstpage}
\pagerange{\pageref{firstpage}--\pageref{lastpage}}
\maketitle

\begin{abstract}
The overwhelming majority of eclipsing white dwarf (WD) binary systems show quasi-periodic variations in eclipse timings on many year timescales. Currently, the mechanism behind these eclipse time variations (ETVs) is not known, with the main competing theories being the planetary hypothesis and the Applegate/Lanza mechanisms. Here, we present a comprehensive study of 43 WD binary systems, the vast majority of which have more than a decade of eclipse timing measurements, analysing their global properties to determine which driving force is the likely origin of the observed ETVs. Long-term, high-speed photometry data obtained with ULTRACAM, ULTRASPEC and HiPERCAM have allowed us to track the evolution of the ETVs in these systems, and analyse any previously unseen trends. From this analysis, we find a clear difference in the level of observed ETVs past the fully convective boundary, where systems with partially radiative companion stars consistently showing high levels of variation. While some systems may be affected by the presence of an unknown planet, the results from this study strongly indicates that an Applegate- or Lanza-like mechanism is the most likely driving force for the timing variations seen in the majority of systems in this sample. However, as found in previous studies, the Applegate/Lanza mechanisms are still not able to reproduce the large and rapid timing variations seen in the vast majority of systems, with the companion star to the WD unable to provide sufficient energy on these short timescales.
\end{abstract}




\section{Introduction}
\label{sec:intro}

White dwarf (WD) binaries contain at least one WD, with the other component typically being either another evolved star, a main sequence star or a sub-stellar object such as a brown dwarf. Eclipsing WD binaries are oriented such that the components of the binary will periodically pass in front of each other from the perspective of an observer. There have been extensive studies of eclipsing WD binaries in particular due to their ability to provide in-depth information on the properties of their components \citep[e.g.][]{Prsa2005,Prsa2022}, assist in calibrating stellar models \citep[e.g.][]{Rebassa-Mansergas2019, Serenelli2021} and test new relationships between properties, such as mass-radius relations in WDs \citep{Parsons2017}. The extensive study of these WD binary eclipses eventually highlighted something unexpected - the eclipses did not always occur at the predicted time \citep[e.g.][]{Lohsen1974}. Eclipse arrival times are expected to be periodic, occurring at predictable intervals once every orbital period.

Eclipse time variations (ETVs) can be caused by the presence of a circumbinary planet, or outer companion in general, due to light travel time (LTT) effects. ETVs are studied through the use of O$-$C (observed eclipse time $-$ calculated eclipse time) plots (see Section \ref{sec:o-c}), the specific shape of which can reveal a lot about the origin of the ETVs. For example, a decrease in the orbital period of the binary will lead to a downward parabola shaped O$-$C diagram, as each eclipse comes earlier compared to a linear ephemeris \citep[e.g.][]{Hermes2012}. In systems with a circumbinary planet, the O$-$C diagram is some form of quasi-sinusoidal variation, as the centre of mass of the binary moves closer to and further from the observer. While many O$-$C diagrams of WD binary systems show quasi-sinusoidal variation, attempts to fit planetary models have been largely unsuccessful, with none of the models thus far able to predict future ETVs, and many of the predicted orbits being shown to be dynamically unstable (see e.g. \citealt{Horner2013}). Quasi-sinusoidal variations also rule out long-term angular momentum loss mechanisms (such as magnetic braking and gravitational wave radiation) as the primary driver of the ETVs, as the distinguishing feature of these would be that the change would always have a consistent sign (e.g. the period would always decrease). 

The discouraging lack of predictive power of the circumbinary planet models lead to the consideration of alternative explanations for the origin of ETVs. The main alternative to the planetary hypothesis is the Applegate mechanism \citep{Applegate1992}. The Applegate mechanism suggests that as the main-sequence companion in the binary goes through its stellar activity cycle, there are changes in the angular momentum distribution in the star's interior. As the angular momentum is transferred to the star's envelope from the core, the star's oblateness changes \citep{Watson2010}. The change in oblateness changes the stellar quadrupole moment, which then changes the orbital period of the binary via a mechanism described by \citet{Matese1983}. From \citet{Applegate1992}, it then follows that if the activity cycle of the active component of the binary follows some kind of sinusoidal variation, the ETVs would produce a quasi-sinusoidal shape in the O$-$C diagram.

The main problem with the \citet{Applegate1992} mechanism is the energy requirement. Namely, in order to produce the ETVs observed, around an order of magnitude more energy than the entire output of the active component in the same timespan is required. Improvements to the \citet{Applegate1992} model have been made multiple times, in \citet{Brinkworth2006} to include proper integration of the internal structure of the star, in \citet{Navarrete2020} to incorporate more sophisticated magnetohydrodynamical (MHD) models, and in \citet{Navarrete2022} to take centrifugal force into account and its effects on the behaviour of the star's quadrupole moment. However, none of these have fully solved the energy requirement issues with \citet{Applegate1992}.

Alternative mechanisms to drive quadrupole moment fluctuations have since been suggested, the most feasible of which is the work of \citet{Lanza2020}. Initially, \citet{Lanza1998} attempted to generalise the MHD models in \citet{Applegate1992}. \citet{Lanza2005} then redefined how the quadrupole moment is changed within the active companion. Essentially, the Lanza and Applegate mechanisms differ in the way they explain how the quadrupole moment changes. In both mechanisms, the ETVs are a result of the orbital period variations caused by the variation in the stellar quadrupole moment. In Applegate, this is done via the change in oblateness of the active component as a result of the angular momentum transfer. Conversely, Lanza employs localised density changes on the active component through the presence of flux tubes. These flux tubes reduce gas pressure and increase magnetic pressure in that area, reducing the density and allowing material to sink. This change in mass distribution of the star causes the stellar quadrupole moment to change, which then goes on the change the orbital period of the binary. Further work in \citet{Lanza2020} reduces the energy requirements further to two or three magnitudes less than in \citet{Applegate1992}, although it requires a small asynchronisation between the rotation of the active component and the orbital period. Given the extremely short synchronisation timescales of $\sim10^{5}$ years for close WD binaries \citep{Zahn1977} it is unclear if this is possible.

Currently, most studies lean towards an Applegate- or Lanza-like mechanism being the main driver for the observed ETVs in WD binaries. However, an alternative theory attribute the ETVs to a combination of \citet{Applegate1992} and the planetary hypothesis (see e.g. \citealt{Zorotovic2013, Volschow2016}). In this investigation, we analyse the long-term ETVs of a sample of 43 WD binaries with the aim of uncovering the main driving force behind the observed timing variations.

\section{Observations}
\label{sec:obs}

The vast majority of observations in this investigation were taken using ULTRASPEC \citep{Dhillon2014}, ULTRACAM \citep{Dhillon2007} and HiPERCAM \citep{Dhillon2021}; high-speed cameras which take images on timescales of seconds or less, with negligable dead time between exposures. As a result, these cameras can resolve the eclipse ingress and egress features, which typically occur on timescales of the order tens of seconds. In general the more precise the timing, the smaller the mass of the perturbing third object that you can detect, which is useful when considering the planetary hypothesis as the reason for the observed ETVs. For example, with the data from the high speed cameras, one could in principle detect Neptune-mass objects.

Some photometry for RR Cae was obtained at the Sutherland site of the South African Astronomical Observatory (SAAO) using the 1m (Elizabeth) telescope. Before 2020, we used the SAAO STE3 CCD photometer, and after 2020, the Sutherland High-Speed Optical Camera (SHOC; \citealt{Coppejans2013}). Typical exposure times were 12-15 sec and made with no filter to improve the signal-to-noise ratio (S/N) and allow shorter exposure times - necessary given the rapid ingress and egress of eclipses for this system. Reduction of the CCD frames were performed using SAAO software based on the DoPHOT program described by \citet{Schechter1993}. In the case of RR Cae, the small field-of-view of the SAAO cameras meant that only two much fainter stars were available as local comparisons, so that relative photometry has significantly worse S/N. This, and the relative brightness of RR Cae, means that the raw aperture magnitudes for RR Cae were preferred.

\subsection{Targets}
\label{subsec:targets}

The eclipse timing programme for this investigation includes 43 eclipsing WD binaries, made up of 3 double white dwarf (DWD) binaries, 5 white dwarf + brown dwarf (WDBD) binaries, and 35 white dwarf + main sequence binaries (WDMS) - 34 of which are M dwarfs and one K dwarf. Target information can be found in Table \ref{tab:targets}. In WDMS and WDBD binaries, the primary component is always defined as the WD. It is slightly more complicated in DWD systems, however for this investigation we have defined the primary component as the higher mass WD.

Our sample is largely based on the eclipse timing programme presented in \citet{Bours2016}, however we have removed the cataclysmic variables from this study, owing to the difficulty in identifying reliable light-curve features that are always present to time in many of these systems. Our investigation follows on from the work done by \citet{Bours2016}, allowing us to identify any changes that were not predicted or expected in the \textasciitilde 10 years since their work was published.

In our analysis we have used previously published stellar and binary parameters for our system where available (see references in Table \ref{tab:targets}). In the case of systems with no dedicated follow-up observations beyond a discovery light curve or low resolution spectrum, we used the method outlined in \citet{Brown2023} to determine their stellar parameters.

\begingroup
\setlength{\tabcolsep}{4pt}
\renewcommand{\arraystretch}{1.2}
\begin{table*}
    \centering
    \caption{The sample of binaries used in this investigation. The ID relates to the ID number shown in Figure \ref{fig:oc grid}. The identifier corresponds to the SIMBAD name. Type gives whether the binary is a double white dwarf (DWD), white dwarf-brown dwarf (WDBD) or white dwarf-main sequence (WDMS) binary. $T_{0}(\mathrm{BMJD_{TDB}})$ and $P_{\mathrm{orb}}$ represent the best fit linear ephemeris, $M_{\mathrm{pri}}$ is the mass of the primary in solar masses, $M_{\mathrm{sec}}$ is the mass of the companion in solar masses, and $R_{\mathrm{sec}}$ is the radius of the secondary in solar radii. Errors on $M_{\mathrm{pri}}$, $M_{\mathrm{sec}}$ and $R_{\mathrm{sec}}$ are given where possible. Note that the uncertainties on the linear ephemerides are purely statistical. Since many of these systems show significant deviations from a linear epehemeris these should be treated as lower limits. References: (1) \citet{Bours2014} - (2) \citet{Kaplan2014} - (3) \citet{Parsons2020} - (4) \citet{Parsons2012} - (5) \citet{Littlefair2014} - (6) \citet{Parsons2017} - (7) \citet{Parsons2018} - (8) \citet{Parsons2012b} - (9) \citet{Obrien2001} - (10) \citet{Muirhead2022} - (11) \citet{Maxted2004} - (12) \citet{Parsons2010} - (13) \citet{Parsons2013} - (14) \citet{Parsons2017b} - (15) \citet{Pyrzas2009} - (16) \citet{Drake2009} - (17) \citet{Parsons2013b} - (18) \citet{Drake2014} - (19) \citet{Brown2023} - (20) \citet{Parsons2025}}
    \label{tab:targets}
    \begin{tabular}{ccccccccc}
    \hline
    ID & Identifier & Type & $T_{0}(\mathrm{BMJD_{TDB}})$ & $P_{\mathrm{orb}}$ (days) & $M_{\mathrm{pri}}$ ($\mathrm{M}_\odot$) & $M_{\mathrm{sec}}$ ($\mathrm{M}_\odot$) & $R_{\mathrm{sec}}$ ($\mathrm{R}_\odot$) & Ref \\
    \hline
    1 & SDSS J100559.11+224932.3 &   DWD   & 55936.3446722(6) & 0.11601543645(2) & 0.378 $\pm$ 0.023 & 0.316 $\pm$ 0.011 & 0.02066 $\pm$ 0.00042 & 1 \\
    2 & NLTT 11748 & DWD & 55772.041383(4) & 0.2350604842(2) & 0.729 $\pm$ 0.008 & 0.153 $\pm$ 0.007 & 0.0429 $\pm$ 0.0004 & 2 \\
    3 & SDSS J115219.99+024814.4 & DWD & 57460.651024(2) & 0.09986526524(7) & 0.362 $\pm$ 0.014 & 0.325 $\pm$ 0.013 & 0.0191 $\pm$ 0.0004 & 3 \\
    4 & SDSS J085746.18+034255.3 & WDBD & 55552.712765(1) & 0.06509653803(4) & $0.514 \pm 0.049$ & $0.087 \pm 0.012$ & $0.110 \pm 0.004$ & 4 \\
    5 & SDSS J120515.80-024222.6 & WDBD & 57768.0393057(8) & 0.0494652611(3) & 0.39 $\pm$ 0.02 & 0.049 $\pm$ 0.006 & 0.081 $\pm$ 0.006 & 14 \\
    6 & SDSS J141126.20+200911.1 & WDBD & 55991.3887174(9) & 0.08453275006(4) & 0.53 $\pm$ 0.03 & 0.050 $\pm$ 0.002 & 0.072 $\pm$ 0.004 & 5 \\
    7 & SDSS J090812.04+060421.2 & WDdM & 53466.3338(1) & 0.149438082(4) & 0.4756 $\pm$ 0.0036 & 0.340 $\pm$ 0.005 & 0.344 $\pm$ 0.003 & 6,7 \\
    8 & SDSS J220823.66-011534.2 & WDdM & 56175.87953(1) & 0.156505689(1) & 0.416 $\pm$ 0.036 & 0.116 $\pm$ 0.014 & 0.137 $\pm$ 0.011 & 6,7 \\
    9 & SDSS J134841.61+183410.5 & WDdM & 56000.16217(4) & 0.248431699(3) & $0.658 \pm 0.010$ & 0.289 $\pm$ 0.011 & 0.293 $\pm$ 0.009 & 7 \\
    10 & SDSS J083845.86+191416.5 & WDdM & 53469.22005(5) & 0.130112313(1) & 0.4817 $\pm$ 0.0077 & 0.142 $\pm$ 0.013 & 0.183 $\pm$ 0.003 & 6,7 \\
    11 & GK Vir & WDdM & 42543.33745(3) & 0.3443308470(8) & 0.5618 $\pm$ 0.0142 & 0.116 $\pm$ 0.003 & $0.146 \pm 0.003$ & 6,8 \\
    12 & NN Ser & WDdM & 47344.0221(1) & 0.130080185(2) & 0.5354 $\pm$ 0.0117 & 0.111 $\pm$ 0.004 & 0.141 $\pm$ 0.002 & 6,7 \\
    13 & QS Vir & WDdM & 48689.1408(1) & 0.150757508(3) & 0.7816 $\pm$ 0.0130 & 0.382 $\pm$ 0.006 & 0.381 $\pm$ 0.003 & 6,7 \\
    14 & RR Cae & WDdM & 51522.54869(2) & 0.303703650(1) & 0.4475 $\pm$ 0.0015 & 0.169 $\pm$ 0.001 & 0.210 $\pm$ 0.001 & 6,7 \\
    15 & SDSS J010622.99-001456.2 & WDdM & 55059.056100(6) & 0.0850153315(3) & 0.4406 $\pm$ 0.0144 & 0.133 $\pm$ 0.007 & 0.150 $\pm$ 0.002 & 6,7 \\
    16 & SDSS J011009.10+132616.1 & WDdM & 53993.94900(3) & 0.332686782(3) & 0.4656 $\pm$ 0.0091 & 0.179 $\pm$ 0.005 & 0.222 $\pm$ 0.004 & 6,7 \\
    17 & SDSS J031452.12+020607.2 & WDdM & 56195.20623(8) & 0.30529696(1) & 0.5964 $\pm$ 0.0088 & 0.395 $\pm$ 0.012 & 0.377 $\pm$ 0.006 & 6,7 \\
    18 & SDSS J102857.78+093129.8 & WDdM & 56001.0970(3) & 0.23502463(3) & 0.4146 $\pm$ 0.0036 & 0.403 $\pm$ 0.005 & 0.398 $\pm$ 0.003 & 6,7 \\
    19 & SDSS J121010.13+334722.9 & WDdM & 54923.0325(2) & 0.124489874(7) & 0.42 $\pm$ 0.01 & 0.158 $\pm$ 0.006 & 0.200 $\pm$ 0.006 & 6,7 \\
    20 & SDSS J121258.25-012310.2 & WDdM & 54104.20911(7) & 0.335870917(5) & 0.4393 $\pm$ 0.0022 & 0.273 $\pm$ 0.002 & 0.306 $\pm$ 0.007 & 6,7,8 \\
    21 & SDSS J130733.49+215636.7 & WDdM & 56007.2207(1) & 0.216322472(8) & 0.6098 $\pm$ 0.0031 & 0.204 $\pm$ 0.002 & 0.227 $\pm$ 0.007 & 6,7 \\
    22 & SDSS J132925.21+123025.4 & WDdM & 55271.05483(1) & 0.0809662439(3) & 0.3916 $\pm$ 0.0234 & 0.088 $\pm$ 0.004 & 0.121 $\pm$ 0.004 & 6,7 \\
    23 & SDSS J223530.61+142855.0 & WDdM & 55469.0657(2) & 0.14445678(2) & 0.3977 $\pm$ 0.0220 & 0.151 $\pm$ 0.013 & 0.174 $\pm$ 0.004 & 6,7 \\
    24 & V471 Tau & WDdK & 54027.95353(9) & 0.521183462(6) & 0.84 $\pm$ 0.05 & 0.853 $\pm$ 0.029 & 0.816 $\pm$ 0.042 & 9,10 \\
    25 & WD 1333+005 & WDdM & 55611.47668(1) & 0.1219587637(4) & 0.4356 $\pm$ 0.0016 & 0.132 $\pm$ 0.001 & 0.163 $\pm$ 0.003 & 6,7 \\
    26 & RX J2130.6+4710 & WDdM & 52785.1810(5) & 0.52103658(6) & 0.554 $\pm$ 0.017 & 0.555 $\pm$ 0.023 & 0.534 $\pm$ 0.053 & 11 \\
    27 & DE CVn & WDdM & 52784.0548(2) & 0.36413911(1) & $0.51^{+0.06}_{-0.02}$ & 0.41 $\pm$ 0.06 & $0.37^{+0.06}_{-0.007}$ & 12 \\
    28 & SDSS J030308.35+005444.1 & WDdM & 53991.11703(4) & 0.134437690(2) & 0.839 $\pm$ 0.014 & 0.181-0.205 & $0.250^{+0.025}_{-0.024}$ & 13,19 \\
    29 & SDSS J143547.87+373338.5 & WDdM & 54148.2049(1) & 0.125630978(4) & $0.456^{+0.024}_{-0.026}$ & $0.141^{+0.018}_{-0.019}$ & $0.187^{+0.007}_{-0.006}$ & 15,19 \\
    30 & SDSS J142355.06+240924.3 & WDdM & 55648.206150(8) & 0.382004281(1) & $0.365^{+0.066}_{-0.069}$ & $0.434^{+0.101}_{-0.137}$ & $0.401^{+0.051}_{-0.081}$ & 16,19 \\
    31 & SDSS J132518.18+233808.0 & WDdM & 55653.4549(2) & 0.19495883(1) & $0.446^{+0.051}_{-0.042}$ & $0.201^{+0.054}_{-0.082}$ & $0.216^{+0.017}_{-0.033}$ & 16,19 \\
    32 & SDSS J124432.25+101710.8 & WDdM & 53466.35982(6) & 0.227856404(2) & $0.405^{+0.045}_{-0.037}$ & $0.144^{+0.033}_{-0.057}$ & $0.158 \pm 0.011$ & 16,19 \\
    33 & SDSS J093947.95+325807.3 & WDdM & 55587.30899(5) & 0.330989612(4) & $0.505^{+0.024}_{-0.035}$ & $0.173^{+0.047}_{-0.076}$ & $0.191^{+0.007}_{-0.008}$ & 16,19 \\
    34 & SDSS J095719.24+234240.7 & WDdM & 55548.3571(1) & 0.150870802(5) & $0.430^{+0.038}_{-0.031}$ & $0.140^{+0.025}_{-0.047}$ & $0.181^{+0.009}_{-0.011}$ & 16,19 \\
    35 & SDSS J082145.27+455923.4 & WDdM & 55989.03872(2) & 0.509092050(3) & $0.703^{+0.052}_{-0.054}$ & $0.285^{+0.055}_{-0.120}$ & $0.339^{+0.024}_{-0.058}$ & 17,19 \\
    36 & SDSS J092741.73+332959.1 & WDdM & 56074.90612(1) & 2.308225576(7) & $0.526^{+0.074}_{-0.076}$ & $0.628^{+0.043}_{-0.059}$ & $0.824^{+0.058}_{-0.064}$ & 17,19 \\
    37 & SDSS J093508.00+270049.2 & WDdM & 56602.8373(6) & 0.20103381(5) & $0.921^{+0.066}_{-0.053}$ & $0.300 \pm 0.026$ & $0.401^{+0.017}_{-0.012}$ & 18,19 \\
    38 & SDSS J094634.49+203003.3 & WDdM & 56032.94565(8) & 0.252861424(5) & $0.390^{+0.048}_{-0.049}$ & $0.261^{+0.067}_{-0.085}$ & $0.253^{+0.017}_{-0.019}$ & 17,19 \\
    39 & SDSS J095737.59+300136.5 & WDdM & 56014.97507(2) & 1.92612446(1) & $0.977^{+0.121}_{-0.074}$ & $0.636^{+0.063}_{-0.095}$ & $0.670^{+0.069}_{-0.072}$ & 17,19 \\
    40 & SDSS J101356.32+272410.6 & WDdM & 53831.1286(4) & 0.12904024(1) & $0.636^{+0.086}_{-0.070}$ & $0.256^{+0.025}_{-0.031}$ & $0.305 \pm 0.012$ & 17,19 \\
    41 & SDSS J105756.93+130703.5 & WDdM & 56010.0609(5) & 0.12516224(2) & $0.420^{+0.041}_{-0.034}$ & $0.146^{+0.033}_{-0.060}$ & $0.169^{+0.006}_{-0.008}$ & 17,19 \\
    42 & ZTF J123016.59-265551.34 & WDBD  & 60432.0478805(32) & 0.23597766(95) & $0.646 \pm 0.017$ & <0.0211 & $0.126\pm0.005$ & 20 \\
    43 & ZTF J182848.77+230838.0 & WDBD & 60431.1494044(2) & 0.11200670(87) & $0.610\pm0.004$ & $0.0186\pm0.008$ & $0.102\pm0.002$ & 20 \\
    \hline
    \end{tabular}
\end{table*}
\endgroup
\subsection{Reduction and fitting}
\label{subsec:reduction}

Most of the data reduction was performed using the HiPERCAM pipeline \citep{Dhillon2021}, which is equipped to efficiently reduce the data from all three high-speed cameras used to collect our eclipse time data. The initial data reduction is standard aperture photometry, using a reference aperture on a nearby non-variable field object that is brighter than the target. This allows the reduction software to always accurately know the position of the target relative to the reference object, even during eclipse.

We converted all times to BMJD(TDB), which is the modified Julian date of the Solar System barycentre on the barycentric dynamical timescale, using Astropy \citep{astropy:2013, astropy:2018, astropy:2022}. The barycentric correction was chosen over a heliocentric one as the differences between these light travel time corrections can be of the order of a few seconds, which is significantly larger than the precision of the eclipse times measure in this analysis.

Light curve fitting was done using LCURVE, software which models WD binary light curves \citep{Copperwheat2010}. For systems in our sample with past best-fit models using LCURVE, we simply used these models. All parameters were fixed at published values and only the expected mid-eclipse time and the temperature of the secondary star ($\mathrm{T}_{\mathrm{sec}}$) was allowed to vary. Varying $\mathrm{T}_{\mathrm{sec}}$ allows for brightness variations in the secondary, for example due to star spots. For systems without previous LCURVE fits, we use data with the highest SNR and time resolution to estimate the values of the scaled radii of the two components and their temperatures. Once these values are estimated, both scaled radii values and the temperature of the WD are fixed in all future fits. In the majority of cases, the limb darkening of the secondary star has a negligable effect on the eclipse of the WD. The exception is when the secondary contributes a significant amount of flux and is roche distorted. For these systems we use previously published limb darkening coefficients. Occasionally, a slope parameter was used in the model. This was on a case by case basis, as this parameter allows for the correction of any linear trends in the observational data either from external factors such as airmass variations or genuine slope in the data from features in the binary itself (e.g. star spots on the companion). This parameter multiplies the light curve by a factor of \begin{equation}
\label{eqn:slope}
    1 + (x \times \mathrm{s}),
\end{equation}
where $\mathrm{s}$ is the slope value inputted into the LCURVE model file, and $x$ is the time, scaled so it varies from -1 to 1 from start to end of the data.

For each model, the period of the binary is always a fixed value that is taken from the best fit ephemeris, and the initial value for the mid-eclipse time (T0) was obtained from plotting the reduced data and then estimating by eye. 

Once the model is initialised, the data and model are run through two fitting algorithms in order to minimise the residuals and find the best fit values for any model parameters left free to vary. The first algorithm is a simplex algorithm, a simple chi-squared minimisation technique that is useful for getting a rough fit from the initial model. It can get stuck at local minima however, so it is important for this step to have a suitable initial model with an initial value of T0 close to the true value. The model is then fine-tuned with the Levenberg-Marquardt (LevMarq) algorithm \citep{Levenberg1944, Marquardt1963}, which is essentially a combination of the Gauss-Newton and steepest descent methods, and provides a much better fit to the data. LevMarq also estimates uncertainties in the fit, which the simplex algorithm does not do. Once the model has been run through these fitting algorithms, the best fit values (plus errors) for any of the free parameters can be obtained, however for this investigation, only the best fit value of T0 and its error was required for subsequent analysis. 

We do not provide a detailed observing log in this paper due to the huge number of observations, spanning many nights, however a full list of all eclipse times for our systems is available online as supplementary material. The observing conditions can usually be estimated from the final uncertainty in the eclipse time.

\section{The O$-$C Method}
\label{sec:o-c}

The main method of analysing eclipse time variations is through an O$-$C diagram. These show the difference between the observed eclipse time (O) and the predicted (calculated) eclipse time (C) based on a linear ephemeris, which is then plotted against the cycle number of that eclipse. The predicted eclipse time is found using
\begin{equation}
\label{eqn:o-c}
    y(x_{i}) = T_0 + x_{i}P,
\end{equation}
where $y(x_{i})$ is the predicted eclipse time, $T_0$ is a reference mid-eclipse time, $x_{i}$ is the cycle number and $P$ is the orbital period of the binary.

\begin{figure*}
    \includegraphics[width=\textwidth,height=\textheight,keepaspectratio,angle=90,scale=1.255]{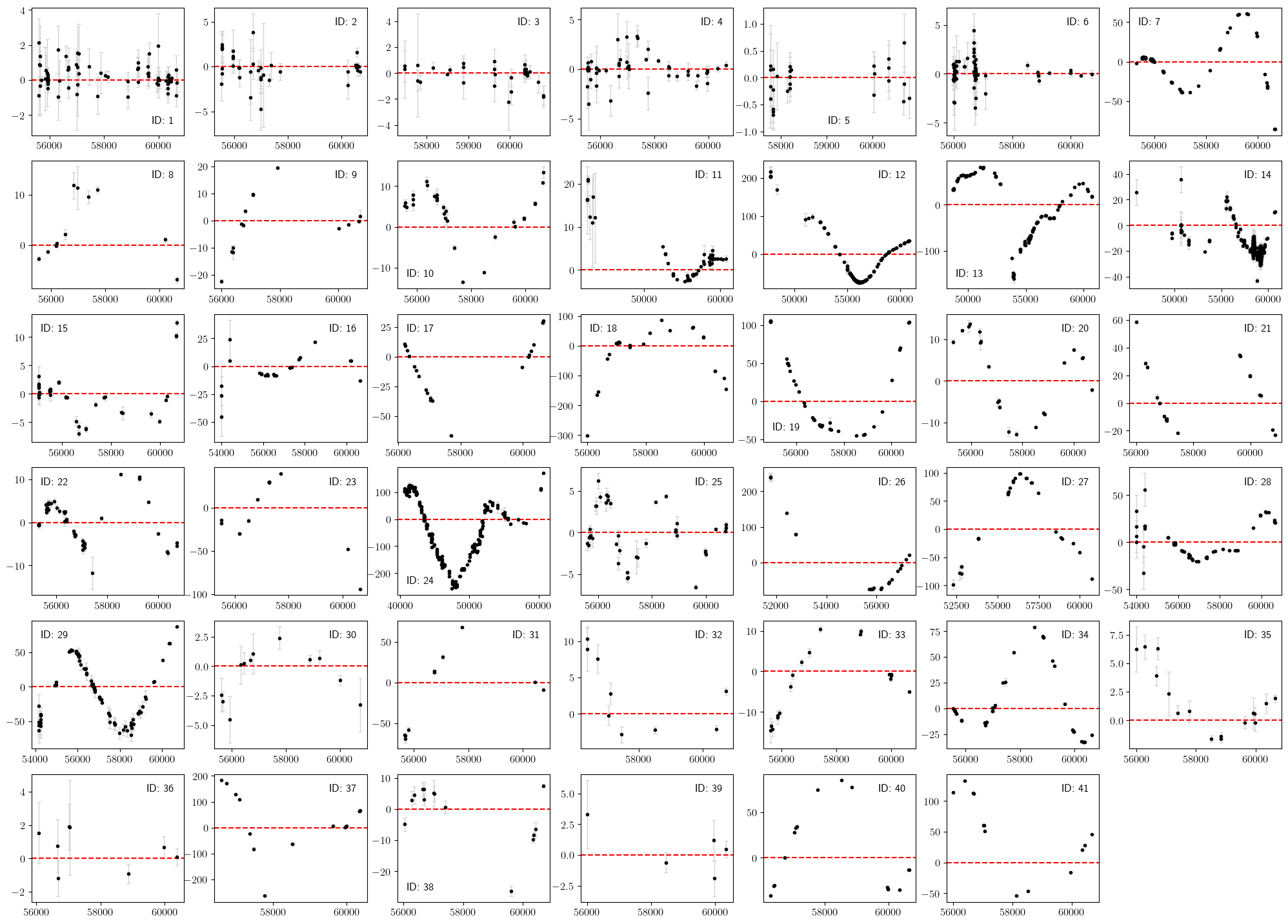}
    \caption{Grid of the O$-$C plots for the WD binaries in the ETV programme with > 2 measured eclipse times. The \textit{y}-axis is the O$-$C value, measured in seconds. The \textit{x}-axis is the MJD(BTDB) time. The red dashed line at \textit{y}=0 is added to highlight the deviations from the expected O$-$C value if no variation in eclipse time was present. The ID number in each plot corresponds to the binary's ID number listed in Table \ref{tab:targets}.}
    \label{fig:oc grid}
\end{figure*}

If there are no variations in the eclipse times, i.e. the eclipses occur exactly when they are predicted, then the O$-$C plot will be a horizontal line centred at 0. As in \citet{Bours2016}, we assume a best fit linear ephemeris for each binary, which is updated with the addition of each new observation. As can be seen in Figure \ref{fig:oc grid}, the vast majority of systems in the sample do not show linear variation, so each new observation can alter the best fit ephemerides significantly enough that it is important to take into account. The current best fit ephemerides for each object can be found in Table \ref{tab:targets}, the errors on which are statistical errors based on the assumption of a linear ephemeris for each system.

Figure \ref{fig:oc grid} shows the O$-$C diagrams for each of the target binaries in our sample that have five or more observations at the time of writing. It can be seen that the O$-$C variations in these binaries are quite different, allowing us to conclude that the driving force behind ETVs does not produce a uniform effect.

\section{Trends and global properties}
\label{sec:trends}

While O$-$C diagrams are useful for telling us about the variations in individual systems, in order to study the global properties of the ETVs in WD binaries, we study how the RMS (root mean square) changes as a function of different binary parameters. The RMS is simply a measure of the amount of variation there is between the measured T0 and the expected best fit value. This method reveals any previously unknown trends in the ETVs as a function of the binary parameters, such as the separation or masses of the components, for example. The RMS of the ETVs for each system is calculated using 
\begin{equation}
\label{eqn:rms}
    \mathrm{RMS} = \sqrt{\frac{1}{N}\sum_{i}\left(\frac{y_{i} - y(x_{i})}{\sigma_{i}}\right)^{2}},
\end{equation}
where $y_{i}$ is observed eclipse time, $y(x_{i})$ is the calculated eclipse time, $\sigma_{i}$ is the uncertainty in the eclipse time, and $N$ represents the number of data points. The RMS is calculated using the best fit linear ephemeris (shown in Table \ref{tab:targets}), with the resulting O$-$C values shifted to have a mean of zero. 

\subsection{Baseline}
\label{subsec:baseline}

\begin{figure*}
    \includegraphics[width=\textwidth]{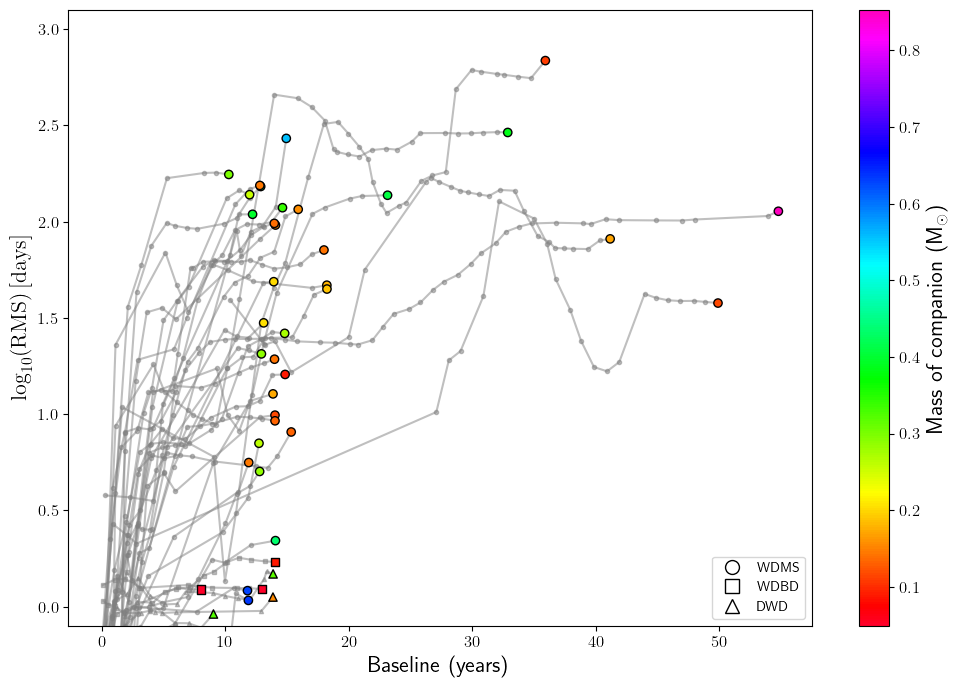}
    \caption{The log of the RMS of the O$-$C values as a function of the baseline of observation, measured in years. The shape of the points indicates the type of binary - a circle means the binary is WDMS, a square means WDBD, and a triangle indicates a DWD binary. The colour of the point indicates the mass of the companion following the colour bar on the right side of the plot. The grey tracks show the evolution of the RMS with baseline for each point.}
    \label{fig:baseline}
\end{figure*}

The baseline of observations for each binary is simply the length of time between the first and last eclipse timing measurements. While not a feature of the binaries themselves, studying the ETVs as a function of baseline can still give useful insights. For example, it could highlight the need for a minimum period of observation before robust conclusions can be drawn if the RMS levels out after a certain baseline. Figure \ref{fig:baseline} shows how the RMS evolves as the baseline increases for each binary in our sample, allowing us to study whether there are any general trends in RMS evolution that could affect any conclusions drawn from the sample.

All but three of the baselines are larger than 10 years, with a typical baseline of roughly 15 years. Figure \ref{fig:baseline} is similar to Figure 4 in \citet{Bours2016}, with mostly the same sample of binaries, just longer baselines. One of the conclusions in \citet{Bours2016} was that it would take at least 10 years for the true RMS of the variations in WD binaries to be seen. They then go on to predict that the RMS of the systems in their sample with baselines less than 10 years would likely increase to match the higher RMS values found in those systems with longer baselines of several decades. This is clearly not always the case, which can be seen in Figure \ref{fig:baseline} as the majority of systems still show lower RMS values than several of the systems with many decades of timing data. For the first time we also see evidence for a significant number of WD binaries that do not show large timing variations, even with a long baseline of observations. Another point to note is that the RMS values of many systems have continued to change after 10 years. This again goes against the conclusions from \citet{Bours2016}, which expected the RMS of the variations to level out after a baseline of 10 years. It is also worth noting that while most systems show large timing variations within a few years - increasing rapidly at first then roughly levelling off, some have slowly rising RMS values that gradually increase with the baseline. The baselines for the DWD binaries are still sufficiently short that we do not expect to see any angular momentum loss from gravitational wave radiation, as we would expect this to be negligible on these timescales given the systems long orbital periods and low WD masses.

Figure \ref{fig:baseline} shows two WDMS binaries with consistently low RMS values on the order of $\mathrm{log}_{10}(\mathrm{RMS}) \approx 0.1$. These binaries both have periods of greater than one day, and are the only two binaries in our sample with periods this long, with most of our systems having periods in the range 0.05 - 0.5 days. The longer period implies both stars in the binary are further from the centre of mass, meaning that any quadrupole moment variations need to be considerably larger in order to induce significant orbital period variations. Hence, the low RMS values of these two longer period systems would appear to be consistent with the reduced impact of an Applegate-like mechanism at wider separations. These binaries also host some of the most massive secondary stars in our sample, which is a significant confounding factor (see Section \ref{subsec:M2} for a discussion of how the companion mass affects the observed ETVs).


In the same group of points that show very little variation in their O$-$C values are the DWD and WDBD binaries. If an Applegate-type mechanism is the driving force behind the observed ETVs then this makes sense, as this mechanism should not be present in the DWD systems and should be extremely weak in WDBD binaries due to: 1) the low luminosity of BDs (meaning that they do not have a lot of energy available to drive orbital period variations) and; 2) the low masses and small radii of BDs, meaning that a considerable change in the quadrupole moment is required to have any impact on the orbit. It is interesting to note, however, that one of the WDBD binaries has a slightly higher RMS value than the rest, indicating the presence of some low level ETV. This can be seen in the O$-$C diagram for object ID: 4 (SDSS J0857+0342) in Figure \ref{fig:oc grid}. The companion in this system is close to the substellar boundary (see \citealt{Dieterich2018} for a discussion on this topic), and may in fact be a very low mass star, rather than a brown dwarf, which would then explain the low level ETVs observed.


The last point within the group of binaries that have $\mathrm{log}_{10}(\mathrm{RMS})$ values of less than 0.5 is a WDMS binary. This point, corresponding to SDSS J1423+2409, shows the most variation out of the points in this group, but still significantly less than the majority of WDMS binaries in our sample (and the small variations in the other WDMS binaries have already been explained by the larger separation). SDSS J1423+2409 is the lowest WD mass WDMS binary in the sample, with the WD mass similar to those in the DWD binaries. The companion mass is close to that of a star on the fully convective boundary \citep{Chiti2024}. It is unclear why either of these features would cause the variations in this system to be so low, but it is interesting to note the physical differences in this systems compared to the others. SDSS J1423+2409 also lacks any detailed study, therefore the stellar parameters may be inaccurate and the companion may well be significantly less massive than previously suggested, which would make this system more consistent with those that show similar RMS values.

\subsection{Effects of the companion's interior structure}
\label{subsec:M2}

\begin{figure*}
    \includegraphics[width=\textwidth]{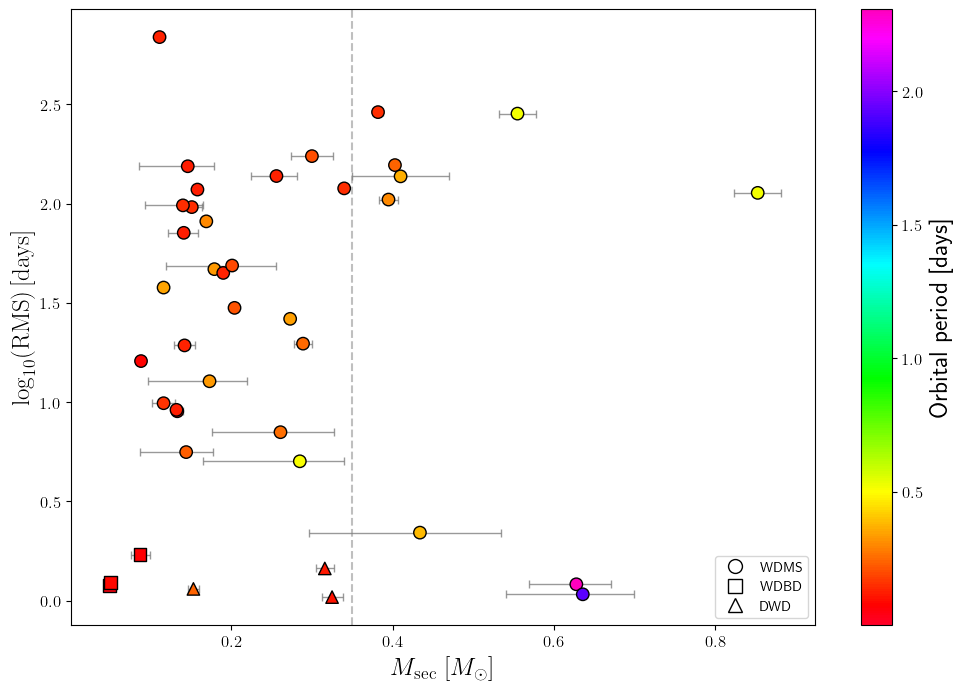}
    \caption{The log of the RMS of the O$-$C values as a function of the mass of the companion ($M_{\mathrm{sec}}$) measured in solar masses. The shape of the points indicates the type of binary, with circles representing WDMS binaries, squares representing WDBD binaries, and triangles representing DWD binaries. The colour of the points represents the orbital period following the colour bar on the right of the plot. Errors on the \textit{x}-axis are shown in grey. The vertical grey, dotted line represents the convective boundary at 0.35$M_{\odot}$.}
    \label{fig:M2}
\end{figure*}

Figure \ref{fig:M2} shows how the RMS of the ETVs is related to the mass of the companion ($M_{\mathrm{sec}}$) in the binary. If the DWD binaries are disregarded (as they show no variation), and excluding the two systems with periods longer than a day, then a trend can be observed between the two parameters, where the RMS increases as $M_{\mathrm{sec}}$ increases. This is particularly interesting when considering the argument between the planetary hypothesis and the Applegate mechanism as the driving force for these ETVs. If the planetary hypothesis is the main mechanism causing the observed variations, then we would expect no trend between the RMS and the mass of the companion. This is because these relatively small changes in $M_{\mathrm{sec}}$ should have no bearing on whether a planet is present or not in the system. Meanwhile, the Applegate mechanism does predict this effect. The fractional energy required to drive a binary's observed O$-$C variations (${\Delta E}/{E_{\mathrm{sec}}}$) through the Applegate mechanism is given by \begin{equation}
\label{eqn:applegate}
    \frac{\Delta E}{E_{\mathrm{sec}}} \propto a^{2} M^{-3.45}_{\mathrm{sec}},
\end{equation} 
 \citep{Volschow2016}, where $a_{bin}$ refers to the binary separation, and $M_{\mathrm{sec}}$ is the mass of the companion.

$M_{\mathrm{sec}}$ and the energy required to drive the variations are inversely proportional, therefore, as $M_{\mathrm{sec}}$ increases, the energy required as a fraction of the total energy available to the star decreases. This makes sense, as more massive stars are more luminous and therefore have more energy available to them to drive quadrupole moment fluctuations. From this, it can be concluded that systems with larger mass companions will show larger variations, as a lower fraction of their total energy is required to drive them via the Applegate mechanism.

Another interesting point is the clear change in behaviour at the fully convective boundary of 0.35$M_{\odot}$ \citep{Chiti2024}. Above this boundary, almost everything shows large variations (with three notable exceptions - SDSS J0927+3329, SDSS J0957+3001, and SDSS J1423+2409), while below it there is a wide range from almost no variation to that of the same level as above the convective boundary. This could support a mechanism such as the Applegate mechanism being the driving force behind the ETVs, as such mechanisms are very reliant on interior stellar structure. Around the fully convective boundary, there is an observed change in the complexity of a star's magnetic field \citep{Kochukhov2021}, where stars with masses greater than $\sim 0.4 \mathrm{M}_{\odot}$ are only observed to have multipolar fields. Below this boundary, there is a dichotomy, with fully convective stars showing a mix of strong, asymmetric, dipolar fields and weak, non-dipolar, asymmetric fields. It may be that more complex fields are less efficient at transporting angular momentum, thus explaining the change in behaviour at the fully convective boundary, if the Applegate mechanism is indeed the driving force behind the ETVs. The large spread in RMS values for binaries with fully convective secondaries seen in Fig. \ref{fig:M2} can also be explained by the clear bimodality in the field geometries of fully convective stars \citet{Kochukhov2021} discusses if they show a similar spread in field geometries.

\begin{figure*}
    \includegraphics[width=\textwidth]{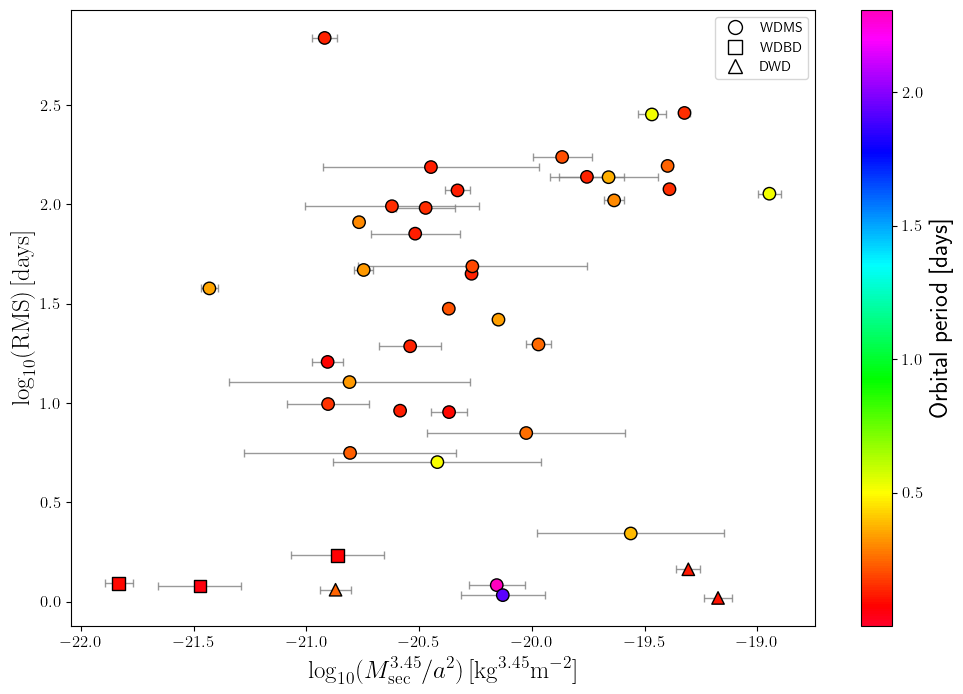}
    \caption{The log of the RMS of the O$-$C values as a function of log($M^{3.45}/a^{2}$). The shape of the points indicates the type of binary, with circles representing WDMS binaries, squares representing WDBD binaries, and triangles representing DWD binaries. The colour of the points represents the orbital period following the colour bar on the right of the plot, and the size of the points represents the mass of the companion, with larger companion masses shown as larger points. Errors on the \textit{x}-axis are shown in grey. Smaller values on the \textit{x}-axis mean a larger fraction of the available energy is needed to drive ETVs via the Applegate mechanism.}
    \label{fig:M2/a}
\end{figure*}

There are three points that do not appear to follow the trends in Figure \ref{fig:M2} discussed above. The two points in the lower right, with values of $M_{\mathrm{sec}}$ around 0.63 $\mathrm{M}_{\odot}$ correspond to the WDMS binaries SDSS J0927+3329 and SDSS J0957+3001, which are the only two binaries in our sample that have periods of greater than one day. Due to observing schedule constraints, the longer period binaries are harder to observe as it is less likely that they will be eclipsing whilst we are able to observe them. Therefore, the number of eclipse times we have measured for these systems is significantly smaller than for other WDMS systems. Nevertheless, it is clear that both of these systems show very little timing variations to date. The $a^{2}$ term in Eqn. \ref{eqn:applegate} indicates that systems require a larger fraction of the total energy to drive the ETVs at wider separations (longer periods), so we would expect these systems to show very little O$-$C variations. Figure \ref{fig:M2/a} shows the ETVs as a function of the energy requirements for the Applegate mechanism, following Eqn. \ref{eqn:applegate}, where we have inverted the quantities in this equation such that smaller values on the \textit{x}-axis of Fig. \ref{fig:M2/a} correspond to a larger fraction of the available energy being required. With this parameter, it is much easier to see that the two long period binaries do follow the same trend as the rest of the WDMS binaries when separation is taken into account.

The third system that does not seem to follow the same trend as the rest of the WDMS binaries is SDSS J1423+2409. This system appears to show variations that are much smaller than we would expect from a binary with $M_{\mathrm{sec}}=0.434^{+0.101}_{-0.137} \,\mathrm{M}_{\odot}$. However, taking into account the relatively large errors on this value, then the system follows the same trend as the others if $M_{\mathrm{sec}}$ lies at the lower end of its error range.

The system showing the highest variation, with $\mathrm{log}_{10}(\mathrm{RMS}) = 2.8$ days, is NN Ser, which has a low mass secondary star ($M_{\mathrm{sec}}=0.111 \pm 0.004\,\mathrm{M}_{\odot}$). Despite this, the system still shows larger ETVs than any other in our sample. Part of this may be explained by the fact that the baseline of observations for NN Ser is much longer than most, and the timing observations are more frequent, however it could be indicative of a secondary driving force on top of the Applegate mechanism. For example, if a circumbinary planet was present in the NN Ser system, then the LTT effect would add to the ETVs from the Applegate mechanism, thus increasing the observed ETVs for the system.

\begin{figure*}
    \includegraphics[width=\textwidth]{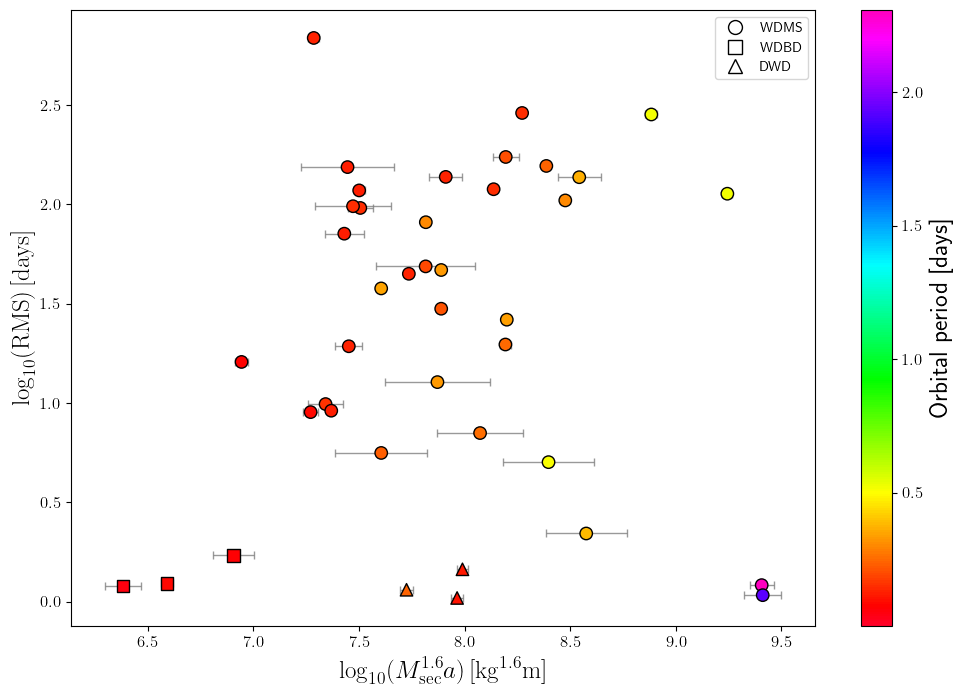}
    \caption{The log of the RMS of the O$-$C values as a function of log($M^{1.6}a$). The shape of the points indicates the type of binary, with circles representing WDMS binaries, squares representing WDBD binaries, and triangles representing DWD binaries. The colour of the points represents the orbital period following the colour bar on the right of the plot, and the size of the points represents the mass of the companion, with larger companion masses shown as larger points. Errors on the \textit{x}-axis are shown in grey. Smaller values on the \textit{x}-axis mean a larger fraction of the available energy is needed to drive ETVs via the Lanza mechanism.}
    \label{fig:M2a}
\end{figure*}

The \textit{x}-axis in Fig. \ref{fig:M2/a} follows the proportionality in the Applegate mechanism given in Eqn. \ref{eqn:applegate}. Following the \citet{Lanza2020} mechanism, we can find an equivalent proportionality 
\begin{equation}
\label{eqn:lanza}
    \frac{\Delta E_{\mathrm{rot}}}{E_{\mathrm{sec}}} \propto a^{-1}_{\mathrm{bin}}M^{-1.6}_{\mathrm{sec}}.
\end{equation}
Figure \ref{fig:M2a} shows the variability as a function of $aM^{1.6}$ in a similar way to Fig. \ref{fig:M2/a}. The main result from this plot is that, other than the two long period systems which show little variability, the trend broadly agrees with that seen in Fig. \ref{fig:M2/a}. Systems towards the left-hand side require a higher fraction of the total energy to be used to drive variations, so we might expect them to show smaller variations, while those towards the right-hand side of the plot need to use a much smaller fraction of their total energy to drive the observed ETVs, so we might expect to see larger variations here. However, Fig. \ref{fig:M2a} shows that we actually see lower variability (albeit with only a couple of systems). The trends in Figures \ref{fig:M2/a} and \ref{fig:M2a} appear to tentatively agree with Applegate-like over Lanza-like mechanisms as the main driving force for the observed ETVs, however more data are needed, particularly for systems with higher mass secondaries and long orbital periods.

\subsection{Estimated energy requirements}
\label{subsec:pmod}

With our results indicating either the Applegate mechanism or the Lanza mechanism as the origin for the observed ETVs in our binary sample, it is important to check the energetic feasibility of both of these mechanisms. Previous studies have found that very few of the eclipsing WD binaries with observed ETVs are energetically feasible with either mechanism, with the energy required to drive the observed ETVs being larger than the energy output of the secondary star as a whole (see e.g. \citealt{Pulley2025}). The fractional energy requirement to drive the observed variations for the Applegate mechanism (${\Delta E}/{E_{\mathrm{sec}}}$) is given by 
\begin{equation}
\label{eqn:applegate energy feasibility}
    \frac{\Delta E}{E_{\mathrm{sec}}} = k_{1} \cdot \frac{M_{\mathrm{sec}}R^{2}_{\mathrm{sec}}}{P^{2}_{\mathrm{bin}}P_{\mathrm{mod}}L_{\mathrm{sec}}} \cdot \left(1 \pm \sqrt{1 - k_{2}G\frac{a^{2}M_{\mathrm{sec}}P^{2}_{\mathrm{bin}}}{R^{5}_{\mathrm{sec}}}\frac{\Delta P}{P_{\mathrm{bin}}}} \right)^{2} ,
\end{equation}
\citep{Volschow2016}, where $M_{\mathrm{sec}}$, $R_{\mathrm{sec}}$ and $L_{\mathrm{sec}}$ are the companion mass, radius and luminosity respectively, $P_{\mathrm{bin}}$ and $a$ are the period and separation of the binary respectively, $P_{\mathrm{mod}}$ is the period of the observed ETVs, $\Delta P / P_{\mathrm{bin}}$ is the fractional period change of the binary, equal to $4\pi K/P_{\mathrm{mod}}$ where $K$ is the semi-major amplitude of the ETVs, assuming they vary in a roughly sinusoidal manner. $G$ is the gravitational constant, $k_{1}$ = 0.133 and $k_{2}$ = 3.42 (constants for low mass stars, as defined in \citealt{Volschow2016}). The fractional energy requirement for the \citet{Lanza2020} model is
\begin{equation}
\label{eqn:lanza energy feasibility}
    \frac{\Delta E_{\mathrm{rot}}}{E_{\mathrm{sec}}} = \frac{1}{3}\frac{ma^{2}_{\mathrm{bin}}}{L_{\mathrm{sec}}P_{\mathrm{mod}}}\left(\frac{\Delta P}{P_{\mathrm{bin}}}\right)\left(\frac{2\pi}{P_{\mathrm{bin}}}\right)^{2},
\end{equation}
where $\Delta E/{E_{\mathrm{rot}}}$ is the fractional energy required to drive the observed ETVs and $m$ is the reduced mass of the binary, defined as $M_{1}M_{\mathrm{sec}}/(M_{1}+M_{\mathrm{sec}})$, with the other parameters defined the same as in Equation \ref{eqn:applegate energy feasibility}.

Most of the parameters in both of these equations were taken from literature, with $M_{\mathrm{sec}}$, $R_{\mathrm{sec}}$ and $P_{\mathrm{orb}}$ listed in Table \ref{tab:targets}. $L_{\mathrm{sec}}$ was calculated using a standard mass-luminosity relation for low mass stars - $L_{\mathrm{sec}} \propto M^{2.6}$ \citep{Volschow2016}. $P_{\mathrm{mod}}$ was found using a mix of autocorrelation \citep[see e.g.][]{Hassani2024} and Lomb-Scargle periodogram (\citealt[][]{Lomb1976, Scargle1982}) analysis of the O$-$C curves of each system. From Figure \ref{fig:oc grid}, it is clear that none of our targets show unambiguous repeating behaviour that would allow for a period of variation ($P_{\mathrm{mod}}$) to be accurately determined. As such, the values of $P_{\mathrm{mod}}$ used for this analysis are fairly uncertain, and in many cases, we are only able to estimate a minimum value for $P_{\mathrm{mod}}$ based on the currently available data (see Appendix \ref{app:energy} for further information).

Out of our sample, only QS Vir showed two clear signals in either periodogram. We therefore calculated the energetic feasibility for both of these values of $P_{\mathrm{mod}}$, labelled as QS Vir (long) and QS Vir (short). Interestingly NN Ser, which previously has shown evidence of multiple period signals in its O$-$C variations \citep{Marsh2014}, did not show a significant second peak in either periodogram, with only the longer of the two previously suggested periods showing significantly.

\begin{figure*}
    \includegraphics[width=0.8\textwidth]{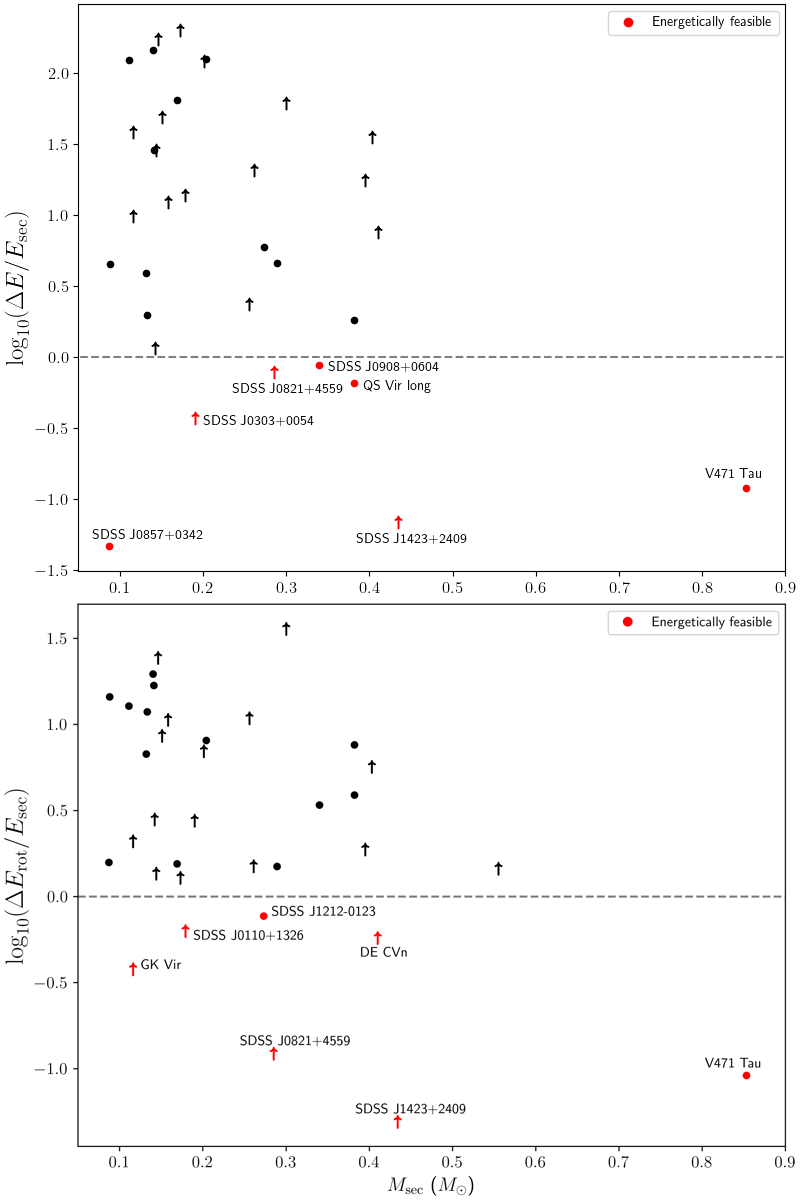}
    \caption{The energetic feasibility of the Applegate mechanism (top) and the Lanza mechanism (bottom) for the binaries in our sample, showing log of the fractional energy requirement versus mass of the companion in solar masses. The grey dashed line represents the feasibility criterion, and the red points are the systems identified as being energetically feasible, labelled with the binary name. Points represented by an arrow represent systems for which the fractional energy requirement is only a minimum value, as shown in Appendix \ref{app:energy}.}
    \label{fig:applegate}
\end{figure*}

Figure \ref{fig:applegate} shows the fractional energy required to drive the observed ETVs for our sample following the Applegate mechanism (Equation \ref{eqn:applegate energy feasibility}, top) and the Lanza mechanism (Eqn. \ref{eqn:lanza energy feasibility}, bottom). The grey dashed line represents a fractional energy of 1, and so above this the energy required to drive the observed ETV's is larger than the total energy output of the active component. Figure \ref{fig:applegate} shows that only a small proportion of the binary systems in our sample fall below this line for both mechanisms, indicating that there are only a few potentially energetically feasible systems in our sample. It should be noted that in this analysis, we define a system as being energetically feasible if their fractional energy is below one, i.e. if the energy required to drive the observed ETVs is less than the total energy output of the binary. In reality, it is likely unrealistic to have a system where the ETVs require energy equal to the companion star's energy output. However, given the fact that only a small proportion of systems are defined as energetically feasible even under the generous definition used in this analysis, this is not seen as an issue as of now. In future, when a model exists that allows the majority of systems to be shown as energetically feasible, it will be important to properly define what proportion of the binary's total energy output can be used to drive ETVs. It is also important to note that the value of $P_{\mathrm{mod}}$ for each binary is based on the dominant signal in the O$-$C plot. Large, rapid timing variations (such as those seen in QS Vir) would require a significantly higher fraction of the binary's energy output.

As with previous studies, we find that, in general, the Lanza mechanism is more energetically feasible than the Applegate mechanism by a factor of 10. Interestingly, while three of the seven energetically feasible systems are the same between both the Applegate and Lanza mechanisms, the other systems are different in each.

In both plots in Fig. \ref{fig:applegate}, the feasible systems are split into two groups - those with the log of their fractional energy requirement being significantly below -0.5, and those between 0 and -0.5. Given the definition of feasibility in this investigation, those systems with values higher than -0.5, especially the ones lying very close to the dashed line, are far less likely to be energetically feasible in reality due to the generous assumption of the energy requirements simply being less than the total energy output of the companion star. As such, we will go on to explain the feasibility of the -0.5 to 0 groups (the \say{moderately feasible} groups) more generally, with a more in depth discussion of the < -0.5 group (the \say{easily feasible} groups).

The Applegate \say{moderately feasible} group includes SDSS J0303+0054, SDSS J0821+4559, SDSS J0908+0604 and the longer period signal in QS Vir. Given that the fractional energies of the majority of binaries decrease by a factor of 10 under the Lanza mechanism, it is interesting that almost all the systems in this group show as moderately feasible in Applegate, but not Lanza, with the exception of SDSS J0821+4559 (see later in this Section for a discussion of this system). The rest of this group is likely explained by their binary periods, as these have some of the lower $a_{\mathrm{bin}}$ values in our sample. Binaries with larger separations (and therefore larger $a_{\mathrm{bin}}$ values) are more likely to have lower fractional energy requirements following the Lanza mechanism, while the opposite is true following Applegate, explaining the discrepancy. It is important to remember that the fractional energy value for SDSS J0303+0054 is a minimum value, and so in reality could be quite different.

The Lanza \say{moderately feasible} group is mostly different to the Applegate group, consisting of SDSS J1212-0123, SDSS J0110+1326, DE CVn and GK Vir. These systems have longer periods, which following Eqn. \ref{eqn:lanza} helps to make these systems energetically feasible. Out of these systems, SDSS J0110+1326 and DE CVn both have minimum fractional energy values (see Appendix \ref{app:energy} for more), and so again their actual fractional energy requirements could be quite different.

Figure \ref{fig:applegate} shows SDSS J0857+0342, SDSS J1423+2409 and V471 Tau as being significantly energetically feasible in the Applegate mechanism. V471 Tau can be explained through the high mass of the companion at 0.853 $\mathrm{M}_{\odot}$. Following Eqn. \ref{eqn:applegate}, the energy required to drive ETVs is sensitive to the companion mass, where increasing $M_{\mathrm{sec}}$ decreases the fractional energy requirement. Thus, the high mass companion in V471 Tau works to decrease the energy required to drive the observed ETVs. From the same argument, the separation of V471 Tau, which is higher than most in our sample, is expected to \textit{increase} the energy requirement for Applegate, following Eqn. \ref{eqn:applegate}. However, as $\Delta E/E_{\mathrm{sec}}$ is more sensitive to $M_{\mathrm{sec}}$, and the separation is not much larger than the other binaries in the sample (with a period of half a day), this allows V471 Tau to remain energetically feasible. SDSS J0857+0342 has a small separation, with an orbital period of only 94 minutes. As explained in Section \ref{subsec:baseline}, binaries with small separations will be more sensitive to changes in the quadrupole moment, meaning a smaller change will have a larger effect. Therefore, smaller quadrupole moment fluctuations are needed to produce larger ETVs. Given that smaller quadrupole moment fluctuations require less energy to drive, the system is more likely to be energetically feasible. However, it is important to note that SDSS J0857+0342 is not feasible following the Lanza mechanism, as shown in the bottom plot of Fig. \ref{fig:applegate}. The \citet{Lanza2020} mechanism works via spin-orbit coupling. Given that binary components with smaller separations spin faster, more energy is required to either spin up or spin down the star sufficiently to drive the observed ETVs. Therefore more energy is required to drive the observed ETVs in binaries with small separations, such as SDSS J0857+0342, explaining why it is significantly feasible following the Applegate mechanism, but not Lanza. The reason SDSS J1423+2409 is in the < -0.5 group is less clear, as it has a fairly typical separation and companion mass for our sample. One thing that may factor into SDSS J1423+2409's feasibility is the low RMS values in its O$-$C variations (see Section \ref{subsec:baseline}), as smaller ETVs will require less energy to drive, therefore making it more likely for the binary to be energetically feasible. Given the forms of Equations \ref{eqn:applegate} and \ref{eqn:lanza}, it is likely that the separation (and therefore the period) is responsible for this discrepancy, as it is the biggest difference between the two forms - $a^{2}$ in Applegate as opposed to $a^{-1}$ in Lanza.

V471 Tau and SDSS J1423+2409 are easily feasible (log of their fractional energy requirement < -0.5) following the Lanza mechanism as well as the Applegate mechanism, for the same reasons as explained above. The only other system in the < -0.5 group shown in the bottom plot of Fig. \ref{fig:applegate} is SDSS J0821+4559, which shows small timing variations in its O$-$C plot, and has a slightly longer period at $P_{\mathrm{orb}} = 0.509$ days which aids in its energetic feasibility following Lanza. This system is also feasible following the Applegate mechanism, which is likely a result of the low level ETVs it exhibits, as they will require less energy to drive.

There is one system that appears in the Lanza feasibility plot (bottom of Fig. \ref{fig:applegate}) but not in the Applegate plot. This system is RX J2130.6+4710, and the fact that it is missing in the Applegate energetic feasibility plot is a result of the form of Eqn. \ref{eqn:applegate energy feasibility}, namely the term inside the square root. The properties of RX J2130.6+4710 mean that the square root does not allow for real-valued solutions, and thus the observed ETVs cannot be driven by the Applegate mechanism \citep{Volschow2016}.

The main conclusion from Figure \ref{fig:applegate} is that, while the \citet{Lanza2020} mechanism does provide an overall improvement on the fractional energy values for the binaries (by reducing the fractional energy requirement by a factor of 10), it still cannot explain the large timing variations seen in the vast majority of WD binaries, and further work is required on both mechanisms in order to explain the origin of the observed ETVs in WD binaries.

\section{Conclusions and summary}
\label{sec:conc}

Long-term, regular observations of eclipsing WD binaries over the past decades have highlighted the discrepancy between observed eclipse times and expected eclipse times, and introduced the question of what other processes are taking place in these systems to cause these observed ETVs. The origin of ETVs in WD binaries is not currently well understood, with the general consensus between the main accepted hypothesis switching between two - the planetary hypothesis and the Applegate mechanism multiple times, with new observational data appearing to favour the latter, or at least a variation of it (i.e. the \citealt{Lanza2020} mechanism). Originally, the idea that the ETVs were a result of LTT effects due to unseen extra bodies in the systems, potentially planets, resulted in many potential planetary fits seemingly working well with the data available. However, these models lacked predictability, with new data not matching the future O$-$C variations, causing a surge in interest in improving the Applegate mechanism to solve the energetic feasibility problem.

A comprehensive study of the ETVs of a large sample of eclipsing WD binaries was attempted by \citet{Bours2016}, however the authors mention that many of the systems in their sample do not have a large enough baseline of observations for robust conclusions to be drawn. Here we present an updated overview of a sample of 43 WD binaries with a wide range of system parameters, and study their global properties and trends. Our main conclusions are as follows:

\begin{itemize}
\item As in previous studies, the DWD and WDBD binaries in the sample continue to show essentially no timing variations, with all the WDMS binaries showing some level of variation. This is in agreement with the Applegate/Lanza mechanism rather than the planetary hypothesis as there is no real reason DWD or WDBD systems could not have planets (\citet{Ledda2023} find that planet formation in DWD circumbinary discs is possible).

\item The baseline of observations for the majority of our sample now exceeds 10 years, with the RMS of the variations still appearing to vary significantly past this time. This argues against the conclusions of \citet{Bours2016}, who suggested that this value might level out when a long enough baseline of observations has been obtained. We also found that a significant portion of our sample continue to show relatively low levels of RMS in the ETVs even past this 10 year baseline , suggesting that large ETVs are not inevitable in all WDMS binaries.

\item Analysing the relation between the O$-$C variation and the mass of the companion prefers an Applegate type mechanism, with a clear behaviour change seen at the fully convective limit for MS companions. Binaries with partially radiative companion stars mostly show significant variation (not including those with periods of greater than a day), while fully convective companion systems show a full range of ETVs from very little variation to matching the partially radiative companion systems. This indicates the change in structure of the magnetic field has a profound effect on the ETVs. As the planetary hypothesis has no explanation for this trend, and the Applegate/Lanza mechanisms are based around magnetic activity cycles, this clearly favours the latter mechanism as a driver for the observed ETVs.

\item Finally, our analysis of the energetic feasibility of both the Applegate and Lanza mechanisms failed to reveal any insight into which is preferred, with both showing only seven out of 43 binaries as being energetically feasible. Our results match previous studies in that the Lanza mechanism gives energy requirements that are a factor of ten smaller than the Applegate mechanism.
\end{itemize}

Overall, further work is required to continue to monitor the ETVs and to create a model that fits all of the trends seen here in a way that is energetically feasible. The astrometry data included in the soon-to-be-released \textit{Gaia} data release 4\footnote{\url{https://www.cosmos.esa.int/web/gaia/data-release-4}} should tell us in certain, nearby systems whether a planet is present or not, as the precision should be good enough to show the expected transverse components of proper motion. This has the potential to completely disprove the planetary hypothesis, at least in some systems, which would be a significant indicator that the Applegate mechanism (or potentially something else entirely) is the main driver of ETVs in eclipsing WD binaries.

\section*{Acknowledgements}

AY acknowledges support by the Science and Technology Facilities Council (STFC) (grant ST/Y509541/1).
SGP acknowledges support by STFC (grant ST/B001174/1). 
This paper uses observations made at the South African Astronomical Observatory (SAAO).
IP acknowledges support from the Royal Society through a University Research Fellowship (URF\textbackslash R1\textbackslash 231496).
This project has received funding from the European Research Council under the European Union’s Horizon 2020 research and innovation programme (Grant agreement numbers 101002408 – MOS100PC).
VSD: ULTRACAM, ULTRASPEC and HiPERCAM are supported by STFC grant ST/Z000033/1.

\section*{Data Availability}

Eclipse times for each binary system are available online as supplementary material. Any additional data can be made available upon reasonable request.



\bibliographystyle{mnras}
\bibliography{example} 




\appendix
\section{Energy values}
\label{app:energy}

\begin{table*}
    \centering
    \caption{The binaries in the sample that show variation and their corresponding values used to investigate their energetic feasibility. $P_{\mathrm{mod}}$ represents the period of modulation, i.e. the period of the dominant quasi-sinusoidal variations in the O$-$C plot for that binary, in years. $\Delta E / E_{\mathrm{sec}}$ and $\Delta E_{\mathrm{rot}} / E_{\mathrm{sec}}$ show the fractional energy requirement for the Applegate and Lanza mechanisms respectively, and the 'Min. value?' column indicates whether the value of $P_{\mathrm{mod}}$ was estimated from the baseline of observations in the situation where the baseline was not long enough to see a whole period of the O$-$C variations, thus leading to the fractional energy values being minimum values. RX J2130.6+4710 does not have a $\Delta E / E_{\mathrm{sec}}$ as it's ETVs cannot be driven by the Applegate mechanism (see Section \ref{subsec:pmod}).}
    \label{tab:energy}
    \begin{threeparttable}
    \begin{tabular}{ccccc}
    \hline
    Object & $P_{\mathrm{mod}}$ & $\Delta E / E_{\mathrm{sec}}$ & $\Delta E_{\mathrm{rot}} / E_{\mathrm{sec}}$ & Min. value?\\
    \hline
    SDSS J220823.66-011534.2 & 14.0 & 9.86 & 2.10 & Yes \\
    SDSS J134841.61+183410.5 & 10.5 & 4.62 & 1.50 & No \\
    SDSS J132518.18+233808.0 & 13.9 & 50.60 & 6.99 & Yes \\
    SDSS J124432.25+101710.8 & 11.9 & 11.00 & 1.36 & Yes \\
    SDSS J093947.95+325807.3 & 13.9 & 67.67 & 1.28 & Yes \\
    SDSS J083845.86+191416.5 & 14.0 & 1.17 & 2.81 & Yes \\
    SDSS J095719.24+234240.7 & 11.3 & 237.72 & 19.61 & No \\
    SDSS J142355.06+240924.3 & 14.1 & 0.11 & 0.08 & Yes \\
    SDSS J090812.04+060421.2 & 12.0 & 0.88 & 3.40 & No \\
    DE CVn & 23.1 & 7.62 & 0.58 & Yes \\
    GK Vir & 21.0 & 38.69 & 0.38 & Yes$^*$ \\
    NN Ser & 20.2 & 123.73 & 12.81 & No \\
    QS Vir (long period) & 13.0 & 0.66 & 3.89 & No \\
    QS Vir (short period) & 9.3 & 1.82 & 7.62 & No \\
    RR Cae & 16.4 & 65.03 & 1.55 & No \\
    RX J2130.6+4710 & 15.1 & nan & 1.46 & Yes \\
    SDSS J010622.99-001456.2 & 4.4 & 1.98 & 11.85 & No \\
    SDSS J011009.10+132616.1 & 18.2 & 13.95 & 0.63 & Yes \\
    SDSS J030308.35+005444.1 & 18.2 & 0.82 & 2.76 & Yes \\
    SDSS J031452.12+020607.2 & 12.2 & 17.67 & 1.88 & Yes \\
    SDSS J082145.27+455923.4 & 12.8 & 1.33 & 0.12 & Yes \\
    SDSS J085746.18+034255.3 & 9.0 & 0.05 & 1.58 & No \\
    SDSS J093508.00+270049.2 & 10.3 & 379.29 & 35.94 & Yes \\
    SDSS J094634.49+203003.3 & 12.7 & 5.56 & 1.50 & Yes \\
    SDSS J101356.32+272410.6 & 12.0 & 3.48 & 10.87 & Yes \\
    SDSS J102857.78+093129.8 & 13.0 & 35.40 & 5.67 & Yes \\
    SDSS J105756.93+130703.5 & 12.8 & 104.32 & 24.38 & Yes \\
    SDSS J121010.13+334722.9 & 15.9 & 12.42 & 10.63 & Yes \\
    SDSS J121258.25-012310.2 & 11.0 & 5.93 & 0.77 & No \\
    SDSS J130733.49+215636.7 & 9.6 & 125.08 & 8.08 & No \\
    SDSS J132925.21+123025.4 & 8.1 & 4.53 & 14.47 & No \\
    SDSS J143547.87+373338.5 & 13.2 & 55.63 & 16.85 & No \\
    SDSS J223530.61+142855.0 & 14.1 & 48.79 & 8.59 & Yes \\
    V471 Tau & 33.1 & 0.12 & 0.09 & No \\
    WD 1333+005 & 6.1 & 3.91 & 6.74 & No\\
    \hline
    \end{tabular}
    \begin{tablenotes}
	   \item{ $^*$ Using data only from 2002 onwards.}
    \end{tablenotes}
    \end{threeparttable}
\end{table*}

\bsp	
\label{lastpage}
\end{document}